\documentclass[runningheads]{llncs}
\usepackage[T1]{fontenc}
\usepackage{graphicx}
\usepackage{amsmath}
\usepackage{orcidlink}
\usepackage{caption}
\usepackage{subcaption}
\DeclareCaptionFont{ninept}{\fontsize{9}{11}\selectfont}

\usepackage{bbding}

\begin{document}

\title{Democratizing Foundations of Problem-Solving with AI: A Breadth-First Search Curriculum for Middle School Students}
\titlerunning{Breadth-First Search Curriculum for Middle School Students}

\author{
Griffin Pitts\inst{1}\orcidlink{0009-0004-3111-6118} \and
Kimia Fazeli\inst{1}\orcidlink{0009-0001-7773-7161} \and
Tirth Bhatt\inst{1}\orcidlink{0009-0001-5139-3081} \and
Jennifer Albert\inst{2}\orcidlink{0000-0001-8015-1105} \and
Marnie Hill\inst{1}\orcidlink{0009-0003-4017-8700} \and
Tiffany Barnes\inst{1}\orcidlink{0000-0002-6500-9976} \and
Shiyan Jiang\inst{3}\orcidlink{0000-0003-4781-846X} \and
Bita Akram\inst{1}\orcidlink{0000-0001-5195-5841}
}

\authorrunning{G. Pitts et al.}

\institute{
North Carolina State University, Raleigh, NC, USA \\
\email{\{wgpitts, kfazeli, tjbhatt, mehill6, tmbarnes, bakram\}@ncsu.edu}
\and
The Citadel -- The Military College of South Carolina, Charleston, SC, USA \\
\email{jalbert@citadel.edu}
\and
University of Pennsylvania, Philadelphia, PA, USA \\
\email{jiang33@upenn.edu}
}

\maketitle
\begin{center}
\large\textit{Preprint. Accepted to the 27th International Conference on Artificial Intelligence in Education (AIED 2026).}
\end{center}

\begin{abstract}
As AI becomes more common in students’ everyday experiences, a major challenge for K--12 AI education is designing learning experiences that can be meaningfully integrated into existing subject-area instruction. This paper presents the design and implementation of an AI4K12-aligned curriculum that embeds AI learning goals within a rural middle school science classroom using Breadth-First Search (BFS) as an accessible entry point to AI problem-solving. Through unplugged activities and an interactive simulation environment, students learned BFS as a strategy for exploring networks and identifying shortest paths, then applied it to science contexts involving virus spread and contact tracing. To examine engagement and learning, we analyzed pre- and post-assessments, student work artifacts, and a teacher interview. Results suggest that students engaged productively with the curriculum, improved their understanding of BFS and AI problem-solving, and benefited from learning these ideas within ongoing science instruction. Teacher feedback further indicated that the module fit well within the science curriculum while supporting intended science learning outcomes. We conclude with curriculum and design considerations for broadening access to learning about problem-solving with AI in education.


\end{abstract}

\section{Introduction}
Artificial intelligence (AI) is now present in many of the systems students use in and outside of school \cite{pitts2025student,pitts2026drives}, motivating growing interest in AI literacy for learners \cite{touretzky2019envisioning}. Efforts such as the AI4K12 initiative provide guidance for what understanding AI should entail in primary and secondary education, articulating Five Big Ideas of AI: Perception, Representation and Reasoning, Learning, Natural Interaction, and Societal Impact \cite{touretzky2019envisioning}. Such guidelines for K--12 AI education emphasize that learners should have opportunities to define AI concepts and to develop a conceptual understanding of how AI systems operate and produce outputs \cite{touretzky2019envisioning}. Related initiatives, including MIT’s RAISE, extend this emphasis by calling for technical ideas and social implications to be addressed together \cite{williams2023ai+}.

A practical challenge, however, is ensuring that all students have the opportunity to engage with AI education during classroom instruction. Formal opportunities to learn about AI are limited for students, and when they occur, are often delivered as stand-alone workshops or other add-on experiences that reach a subset of students, rather than being interwoven with content area coursework that draws broad interest from the general population of students \cite{li2024systematic,morales2024unpacking}. Related work suggests that one promising direction is to integrate AI- and computational thinking-related topics into K--12 science instruction in ways that connect to disciplinary content and classroom curriculum, rather than positioning these topics only as stand-alone experiences \cite{boulden2018computational,jiang2022agents}. Further, designing for broad participation means developing learning experiences that are feasible within classroom constraints, reflect teacher and community input, and account for differences in local resources, available instructional time, and teacher support and preparation that can affect curricular accessibility across school contexts \cite{kim2025empowering}. Logistical considerations also intersect with student-level barriers to engaging with AI education. Introductory AI materials reach students with widely different prior experiences, yet many still presume some familiarity with programming or advanced mathematics \cite{morales2024unpacking}. As a result, participation may be narrowed toward students who already feel positively about AI or are more likely to opt in. 

In this work, we broaden access to AI education by developing and evaluating a curriculum that introduces Breadth-First Search (BFS), a classical AI search algorithm, within a middle school science unit on virus transmission. Co-designed with the course teacher to fit existing instructional constraints, the curriculum was implemented as a seven-day module across six middle school science classrooms. It combines guided practice in \textsc{I-SAIL}, an interactive Snap!-based learning environment for teaching AI \cite{akram2022towards,rao2025leveraging}, with unplugged activities connecting BFS to contact tracing and disease spread. Students first traced BFS through map-based pathfinding, then applied it in structured tasks tied to their science curriculum. To examine the feasibility and learning outcomes of this approach, this study addresses two research questions: (RQ1) To what extent does the curriculum support students’ understanding of BFS? and (RQ2) What does teacher feedback suggest about the curriculum’s fit within science instruction and its support for the intended science learning outcomes?

\section{Related Work}

A common entry point to AI in education is for students to interact with simplified AI systems, such as speech recognition tools or basic classifiers, paired with discussion of errors and bias \cite{su2022artificial,yang2022artificial,kim2021and}. One line of work supports this kind of interaction through toolkits that make model-building feasible in classroom settings, including PopBots \cite{williams2019popbots}, where children build, train, and interact with a social robot while engaging with knowledge-based systems, supervised learning, and generative AI concepts. Web-based tools can lower barriers for teachers and learners by making AI interaction possible without extensive setup, including platforms such as Teachable Machine \cite{hollands2024establishing}. Work on rule-based and embodied interaction has also made system logic more visible, such as Calypso for Cozmo \cite{touretzky2017computational}, which combines speech recognition, landmark-based navigation with a visible map, and state-machine programming, as well as tangible systems like Any-Cubes \cite{scheidt2019any} that enable hands-on exploration of AI through physical interaction. These activities can help students recognize that AI systems map inputs to outputs and that their behavior heavily relies on training data and design decisions. However, these interactions typically focus on observable system behavior rather than the procedures or representations that govern how AI systems arrive at those outputs. Motivating the presented study, our work focuses on helping students engage with step-by-step reasoning procedures an AI system can use to represent a problem, explore possible states, and generate a solution.

\section{Curriculum Design}
We designed a curriculum that introduces AI problem-solving through Breadth-First Search (BFS) within middle school science instruction. We selected BFS because it provides a clear, traceable example of the foundations of AI-based problem-solving, where BFS is presented as a procedure that operates over representations of nodes and connections in a network, emphasizing how AI can find shortest paths through repeated exploration steps.

The curriculum was implemented in \textsc{i-SAIL}, an integrated science and AI learning environment built on Snap!, a block-based programming platform extended with custom AI blocks, visual simulations, and scaffolded activities that support AI-focused STEM problem-solving \cite{akram2022towards,rao2025leveraging}. To support students' engagement and intrinsic motivation, the curriculum context was designed to be contextually relevant to students \cite{vaino2012stimulating}. We chose path-finding on a map, using GPS navigation systems that suggest routes and directions as a familiar reference point. The curriculum and learning environment were co-designed with the classroom teacher so that the activities would be feasible within existing instructional time and accessible to students with varied prior computing experience. To support learning while managing cognitive load, the activity sequence used phased scaffolding, with guidance gradually reduced as students gained familiarity with the procedure \cite{fisher2021better}. This design aligns with AI4K12 Big Idea \#2 (Representation and Reasoning), which specifies that K--12 learners should understand how AI systems construct representations of the world and apply procedures to derive new information from what is already known \cite{touretzky2019envisioning}. The module was implemented as a seven-day instructional sequence totaling approximately five hours. It began with an introduction to AI and BFS-related concepts through lecture and guided discussion. Students then completed four simulation-based activities in \textsc{i-SAIL} (Section 3.1), and the remaining sessions extended this work through unplugged activities and science-based applications (Section 3.2).

\subsection{\textsc{I-SAIL} simulation-based learning activities}

\begin{figure}[t]
    \centering
    \begin{subfigure}[b]{0.48\linewidth}
        \centering
        \includegraphics[width=\linewidth]{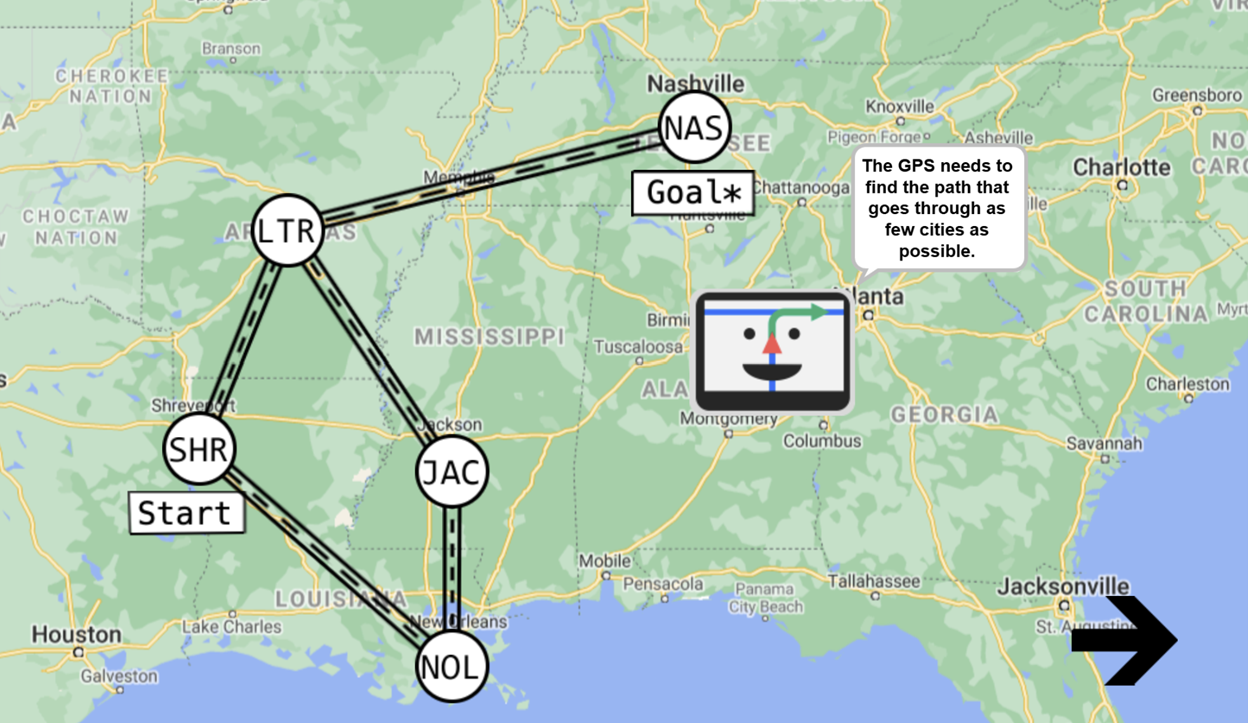}
        \caption{Activity 1}
        \label{fig:placeholder_a}
    \end{subfigure}
    \hfill
    \begin{subfigure}[b]{0.48\linewidth}
        \centering
        \includegraphics[width=\linewidth]{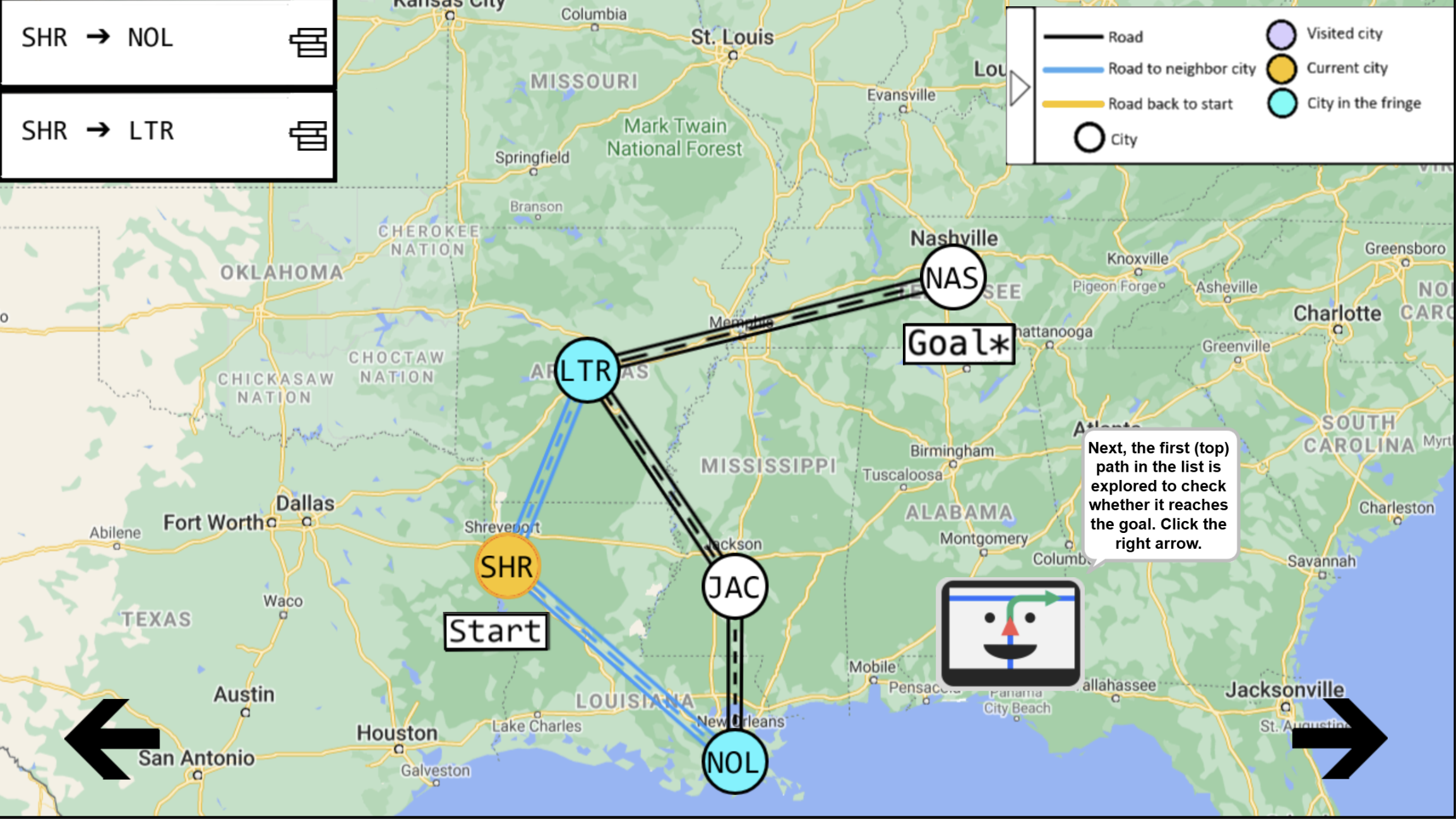}
        \caption{Activity 2}
        \label{fig:placeholder_b}
    \end{subfigure}

    \vspace{0.2em}

    \begin{subfigure}[b]{0.48\linewidth}
        \centering
        \includegraphics[width=\linewidth]{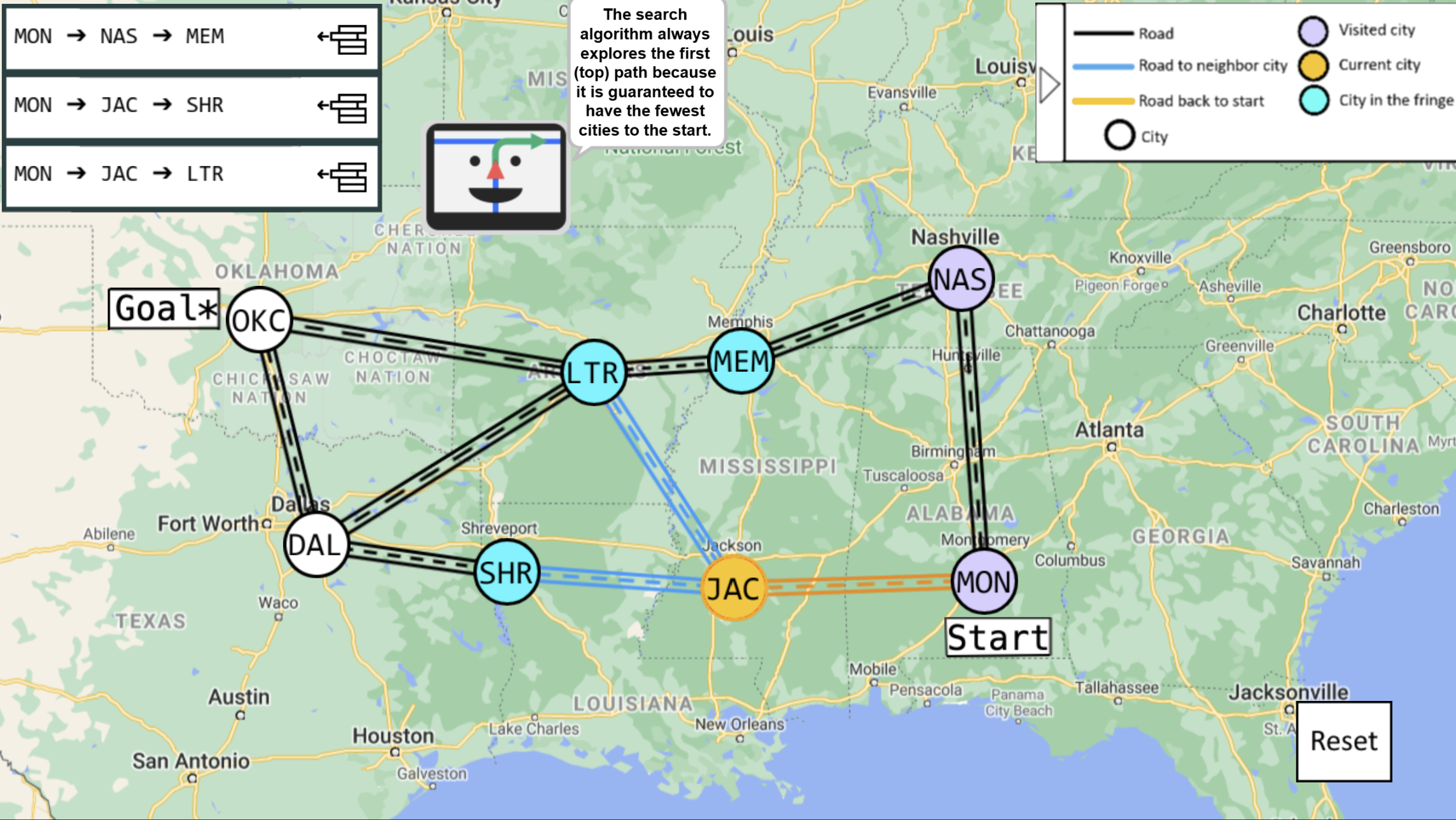}
        \caption{Activity 3}
        \label{fig:placeholder_c}
    \end{subfigure}
    \hfill
    \begin{subfigure}[b]{0.48\linewidth}
        \centering
        \includegraphics[width=\linewidth]{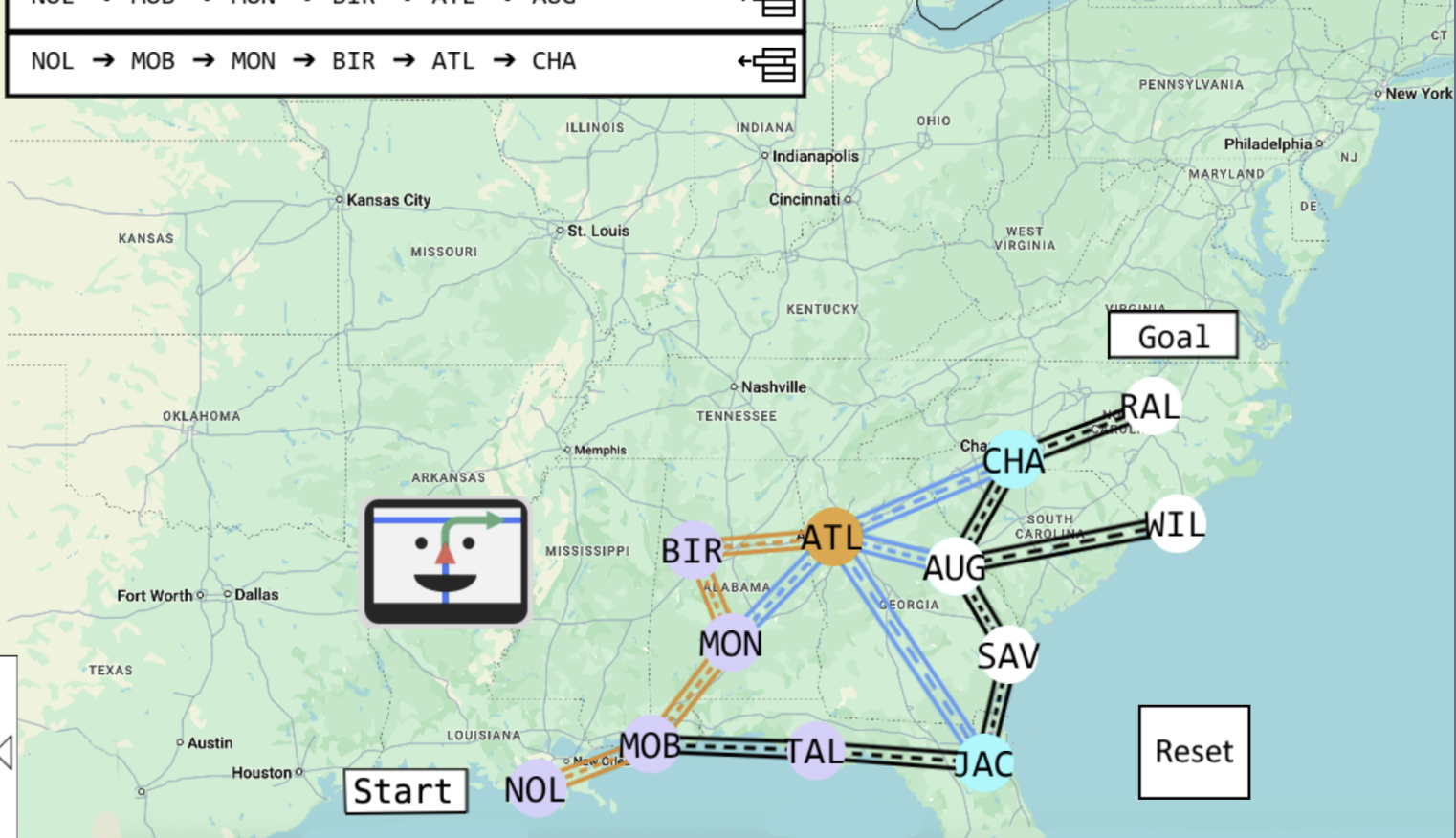}
        \caption{Activity 4}
        \label{fig:placeholder_d}
    \end{subfigure}
    
    \caption{Interface views from the \textsc{I-SAIL} BFS activity sequence}
    \label{fig:placeholder_grid}
\end{figure}

Students learned and practiced BFS through four simulation-based activities in \textsc{I-SAIL}. These activities build on prior iterations of BFS curriculum work \cite{yoder2020gaining,rao2025leveraging,akram2022towards}. In the environment, students worked with map-based graphs in which cities were represented as nodes and roads as edges connecting them (Figure~\ref{fig:placeholder_grid}). A \textit{path} is a sequence of connected cities that represents a possible route between locations. As students traced routes from one city to another, they learned how BFS explores neighboring cities before moving farther outward, while maintaining a \textit{frontier}, or list of discovered but unexplored routes, and a record of \textit{visited} nodes, or cities whose outgoing roads have already been examined.

Activity 1 introduced the need for a path-search procedure through a GPS-style scenario (Figure~\ref{fig:placeholder_a}). Students helped a family plan a trip by finding the shortest route from a start location to a destination. The activity was guided by an on-screen GPS guide, West, who prompted students to propose routes and check whether they had reached the destination in the fewest steps. Activity 2 then introduced BFS as the procedure West uses to find shortest paths (Figure~\ref{fig:placeholder_b}). In this guided activity, students learned the vocabulary and traced the algorithm, focusing on how routes are added to the frontier, removed in queue order, and expanded until the goal is reached. Activity 3 reduced guidance and moved students toward more independent tracing (Figure~\ref{fig:placeholder_c}). Activity 4 extended this work on a larger map, where students again used BFS to find a shortest route, with hints available on request (Figure~\ref{fig:placeholder_d}).

\subsection{Application and supporting activities}

The curriculum also included supporting activities that reinforced students’ understanding of BFS and extended it into their science instruction. This sequence reflected elements of a Use-Modify-Create progression \cite{lee2011computational}: students first applied BFS in structured tracing tasks, then adapted it to contact-network scenarios, and later used it more independently in subject-area problem-solving activities.

Following the four initial BFS activities in \textsc{I-SAIL} on Days 2 and 3, Day 4 shifted to an unplugged, paper-based tracing task that gave students additional practice carrying out the procedure by hand. Day 5 focused on unplugged science activities connecting virus transmission to AI problem-solving over contact networks. In a hands-on \textit{"Glo Germ"} simulation, a small amount of fluorescent powder was placed on a student’s hands to represent an infectious agent. As students interacted, the powder spread through contact and was later revealed under UV light, helping students see how transmission moves through chains of contact. Building on this activity, a card-based task modeled the class as a contact network. Each student received a card, and connections were drawn to represent recent contacts. Green and red dot markers indicated infection states, and students discussed infection likelihood in terms of network distance, or steps, between individuals. This connected the science activity back to BFS as a procedure for tracing shortest contact paths. On Day 6, students used a simulation in \textsc{I-SAIL} to connect BFS reasoning to scientific problem-solving by generating and testing hypotheses about how factors such as temperature and network density affect patterns of virus spread in connected populations. On the final day, the module concluded with a session discussing career pathways, how AI and computing are used in those roles, and their effects on people and communities.

\section{Methods}

The study took place in Fall 2025 in a rural public middle school in the United States (typically ages 11--14). The module was implemented in six health science classrooms across three grade levels: two 6th-grade classrooms, two 7th-grade classrooms, and two 8th-grade classrooms. Each class met in 45-minute class periods, and the module was delivered over seven days by the same instructor. Approximately, 30\% of students at the participating middle school in each grade level were enrolled in the course. In total, 103 students participated as part of regular instruction. This study was approved by North Carolina State University's Institutional Review Board, and parental consent was obtained prior to analysis. The analyses reported include only students with consent and matched data for the relevant measures; after filtering, $N=59$ students with parental consent remained ($n_{6}= 14$, $n_{7}= 29$, $n_{8}= 16$).

Data were collected from three primary sources: (1) pre/post knowledge assessments, (2) student work artifacts, and (3) a teacher interview. Students completed matched BFS knowledge assessments immediately before Activity~1 and after Activity~4 in \textsc{I-SAIL}. The assessment included five multiple-choice items (K1--K5) targeting both general search intuitions and BFS procedural understanding. %Item prompts were: K1, “How does a GPS find the closest path between two cities?”; K2, “You want to find the closest playground from where you are. You can walk along any connected streets. What is the best way to search?”; K3, “How does Breadth-First Search work when finding a path?”; K4, “If Breadth-First Search has already checked all places directly connected to the start, what does it check next?”; and K5, “When is Breadth-First Search the best algorithm to use?” 
Each item was scored for correctness, and the assessment score was computed as the proportion correct across the five items. %Given the post-assessment was administered immediately after the four core BFS activities and prior to later application/reinforcement activities, observed pre- and post- differences primarily reflect learning associated with the activities described in section 3.1. 

%\textbf{Interaction log data.} Alongside the pre-/post-assessment, interaction logs were collected from the \textsc{I-SAIL} learning environment, including timestamps and user actions (e.g., advancing steps, selecting nodes). From this data, we computed summary engagement measures such as time-on-task and interaction counts for each activity. In this paper, we use these logs to characterize how students engaged with the activities and to identify patterns relevant to pacing and curriculum design; in future work, we plan to leverage the same logs for deeper behavioral analyses of students’ strategy use and logical errors during tracing.

Following the learning activities on Day 2 and 3, students completed a paper-based BFS tracing worksheet. Students were given a graph with designated start and goal nodes and were asked to carry out BFS by hand, recording (1) an “explored” list (the nodes they had already checked) and (2) a “path” list (the candidate routes they were considering as the search progressed). This was students’ first time practicing BFS without the environment’s visual supports and served both as reinforcement and as an artifact for examining their procedural understanding of BFS. After the classroom implementation, we conducted a semi-structured interview with the instructor to gather feedback on the curriculum, their experience leading the activities, and opportunities for improvement.

%Students completed brief written reflections in an end-of-module survey on Day 7. The prompts asked students when BFS first made sense to them, which activities supported their learning the most and why, which parts were confusing or made learning difficult and why, what they would change about the activities, and what they would like to learn next. 

%These qualitative data were used to contextualize learning and engagement results and to inform potential revisions to the curriculum.

%On Day 7, students completed brief written reflections about when BFS became clear, which activities helped most, sources of confusion, suggested changes, and future learning interests. After the implementation, we conducted a semi-structured instructor interview on activity flow, student difficulties, classroom experience, and improvement opportunities. These qualitative data contextualized learning and engagement findings and informed potential curriculum revisions.

\section{Results}
We report results in two parts aligned with our research questions. For RQ1, we examine the extent to which the curriculum supported students’ \textbf{understanding of BFS} following the four learning activities through \textsc{I-SAIL}. For RQ2, we summarize findings from the post-implementation \textbf{teacher interview}.

To examine changes in students' understanding of BFS, we analyzed matched pre- and post-assessments administered before and after the four core BFS activities. After filtering for parental consent, and completed pre- and post-assessments, the participant sample included $41$ students. Students demonstrated improved performance from pre- to post-test (pre: $M = 0.683$, $SD = 0.219$; post: $M = 0.790$, $SD = 0.232$). This improvement was statistically significant in both a paired t-test ($t(40) = 2.90$, $p = .006$) and a nonparametric Wilcoxon signed-rank test ($W = 87$, $p = .008$), with a moderate within-subject effect size (Cohen’s $d_z = 0.453$). The largest item-level gain was on K3 ($\Delta = 0.268$); this question focused directly on how BFS works when finding a path, with a correct response describing the search as expanding outward from the start in layers. 

%Figure \ref{fig:prepost} shows students' average pre- and post-assessment scores by classroom grade level per item.

%\begin{figure}[h]
%    \centering
%    \includegraphics[width=.9\linewidth]{fig_prepost.png}
%    \caption{Mean pre- and post-assessment scores by grade and assessment item.}
%    \label{fig:prepost}
%\end{figure}

%\textbf{Insights from clickstream logs.} Interaction logs suggest the BFS curriculum through \textsc{I-SAIL} sustained students’ engagement as the sequence progressed toward more independent work. Time-on-task and interaction volume remained steady and tended to be higher in the later, less-scaffolded activities. In particular, the final activity showed the greatest time-on-task (Activity 4: $\mu = 12.76$ min, $\tilde{x} = 11.01$ min) and the highest interaction volume ($\mu = 190.0$ clicks), compared to Activity 1 ($\mu = 8.39$ min, $\tilde{x} = 7.48$ min; $\mu = 116.2$ clicks).

%The average and median time spent by the students on different activities was as follows: activity 1 ($M = 8.39$ mins, $Me = 7.48$ mins), activity 2 ($M = 9.71$ mins, $Me = 8.83$ mins), activity 3 ($M = 8.08$ mins, $Me = 6.17$ mins), and activity 4 ($M = 12.76$ mins, $Me = 11.01$ mins). 

%The average number of clicks per activity was used as a measure for the interaction that the students had with the activities: activity 1 ($M = 116.2$), activity 2 ($M = 159.9$), activity 3 ($M = 141$) and activity 4 ($M = 190$). 

After working through the \textsc{I-SAIL} learning activities, students completed a paper-based worksheet to practice tracing the BFS procedure without the \textsc{I-SAIL} environment’s supports. Most students (64.5\%) were able to use BFS to accurately identify the shortest path to the ‘goal state’ on the worksheet. Although, even when the final route was correct, students’ written work often omitted parts of the step-by-step record needed to show how the procedure was carried out in full. For instance, students often skipped few documenting intermediate node expansions and the corresponding updates to the list of nodes to explore next, leaving gaps in how they moved through the graph. A second related recurring issue was incomplete or missing records of the exploration order, associated with ordering mistakes.

To better understand how the curriculum was experienced and supported students’ conceptual understanding of AI, we analyzed a semi-structured post-implementation interview with the classroom teacher. The classroom teacher was involved throughout the project, including co-design of the curriculum and learning environment before classroom implementation. Based on the interview analysis, the teacher described the curriculum as fitting well within the health science unit and as supporting students’ understanding of both the science content and the AI ideas introduced in the module. She noted that connecting BFS to virus transmission and contact tracing helped students see the relevance and purpose of the procedure. She also described the learning sequence as effective, explaining that students were first able to “see what the algorithm was doing” through hands-on interaction, while teacher-led discussion helped clarify the “why or the how” behind the process. More broadly, she observed that movement-based and interactive activities supported students’ retention of the material, noting that on days when students were more active they “started to really understand” and “can usually recall what was taught that day too.” The teacher also identified several areas for improvement. She suggested simplifying instructional language and adding more explicit vocabulary instruction for students, and suggested additional teacher-facing supports to better support instructors with limited AI knowledge. It was also noted that students may claim understanding during instruction, while later revealing confusion, indicating a need for more frequent knowledge checks and opportunities for clarification.

%In the post-implementation interview, she provided positive feedback on the AI and BFS curriculum, describing her participation as “one of the coolest things” she had experienced at her school and expressing interest in continued collaboration.

%In relation, she also noted that guided discussion breaks between activities were effective for sustaining students' attention and supporting sense-making, adding that students often "end up learning so much more.. when they start having conversations with one another." 
%Finally, she emphasized that participating in the co-design process and working through the activities in advance helped her feel prepared to teach the module, noting that “if I didn’t have that to begin with, I’m not sure how comfortable I would have been later,” and adding that “the more I can learn about something before I teach it…the better that lesson’s going to go.” 
%Finally, she recommended more explicit comparison across the GPS pathfinding and health science activities to help students recognize that the same procedure applies in both contexts. 

\section{Discussion}

This study examined the implementation of a curriculum designed to broaden access to AI problem-solving by introducing BFS within middle school science instruction. The results and qualitative feedback suggest that this approach supported students’ conceptual understanding of AI and related science learning outcomes. The project was guided by a set of design principles aimed at democratizing access to AI education: (1) making AI education accessible to students without prior computing experience through engaging, well-guided activities, (2) co-designing the curriculum with a classroom teacher to support feasibility and adoption, (3) implementing the curriculum in a rural school context to extend access to AI learning opportunities, and (4) embedding BFS within subject-area instruction to broaden participation.

Student learning gains, including a significant pre-to-post improvement on the BFS knowledge assessment, suggest that middle school students can productively engage with introductory algorithmic reasoning when instruction includes interactive representations and structured guidance. These outcomes were observed in a general elective with no computing prerequisites, suggesting that meaningful engagement was feasible even for students with no programming background. From students’ worksheet traces during the unplugged activity on Day 4, we found that many students could apply BFS to derive the correct shortest path, while still struggling to trace the procedure consistently in a formal manner. Students often shortcut intermediate steps, suggesting incomplete but emerging mental models of the BFS procedure. This raises an instructional question about when curricula should emphasize step-by-step rigor versus allowing intuitive progress. Early correctness may reflect intuitive reasoning that precedes formal procedural alignment, suggesting that insisting on full procedural fidelity too early may suppress sense-making. Future work could examine how and when learners transition from outcome-oriented strategies to more explicit reasoning, and how instruction can leverage partial understandings as stepping stones toward formal algorithmic competence.

The findings also suggest that grounding AI problem-solving in familiar contexts supported students’ understanding of BFS. By introducing the procedure through GPS-style navigation and revisiting it through virus transmission and contact tracing, the curriculum connected an abstract search process to settings students could readily interpret. Teacher feedback further suggests that this subject-area connection helped students see the relevance and purpose of the procedure within the health science unit. Future work could examine how similar subject-area grounded approaches help students connect AI problem-solving to topics across their science curriculum.

\section{Conclusion}
This paper presents the design and classroom implementation of an AI curriculum embedded in middle school science instruction. The paper contributes: (1) a co-designed curriculum, simulation-based activities, and instructional materials that teach BFS by making AI reasoning processes visible and interactive without requiring prior computing experience; (2) a classroom study in a rural middle school health science setting examining students’ learning through assessments, work artifacts, and a teacher interview; and (3) design considerations for broadening access to AI problem-solving in primary and secondary education. These considerations include supporting conceptual accessibility through visual and unplugged representations of AI, partnering with educators to incorporate classroom and community perspectives to curriculum development, and situating AI curriculum in subject-area contexts that help students see AI as something they can understand and participate in, with a stake in how it is used in the future. 

\begin{credits}
\subsubsection{\ackname} This research was supported by the U.S. National Science Foundation (NSF) under Grant \#2405862. Any opinions, findings, and conclusions expressed in this material are those of the authors and do not necessarily reflect views of the NSF.
\end{credits}

\bibliographystyle{splncs04}
\bibliography{ref}
\end{document}